\documentclass[conference]{IEEEtran}

\usepackage{lipsum}
\usepackage{multirow}
\usepackage{subfigure}
\usepackage{graphicx} 
\usepackage[colorlinks, linkcolor=black, citecolor=cyan]{hyperref} 
\usepackage{parselines} 
\DeclareMathAlphabet{\mathbb}{U}{bbold}{m}{n}

\newcommand{\Q}[1]{`#1'}
\usepackage[acronym]{glossaries}
\setacronymstyle{long-short}
\newacronym{AI}{AI}{Artificial Intelligence}
\newacronym{eTSA}{eTSA}{endoscopic TransSphenoidal Approach}
\newacronym{FPS}{FPS}{frames per second}
\newacronym{LSTM}{LSTM}{Long Short-Term Memory}
\newacronym{MAE}{MAE}{Mean Absolute Error}
\newacronym{ML}{ML}{Machine Learning}
\newacronym{RSD}{RSD}{Remaining Surgery Duration}

\newacronym{CRUK}{CRUK}{Cancer Research UK}
\newacronym{DSIT}{DSIT}{Department of Science, Innovation and Technology}
\newacronym{EPSRC}{EPSRC}{Engineering and Physical Sciences Research Council}
\newacronym{IRB}{IRB}{Institutional Review Board}
\newacronym{NHNN}{NHNN}{National Hospital for Neurology and Neurosurgery}
\newacronym{NIHR}{NIHR}{National Institute for Health and Care Research}
\newacronym{UCL}{UCL}{University College London}
\newacronym{WEISS}{WEISS}{Wellcome/EPSRC Centre for Interventional and Surgical Sciences}

\begin{document}

\title{PitRSDNet: Predicting Intra-operative Remaining Surgery Duration in Endoscopic Pituitary Surgery}



\author{\IEEEauthorblockN{Anjana Wijekoon$^{1,2}$, Adrito Das$^{1}$, Roxana R. Herrera$^{1}$, Danyal Z. Khan$^{1,3}$, John Hanrahan$^{1,3}$, Eleanor Carter$^{3}$,\\
Valpuri Luoma$^{3}$, Danail Stoyanov$^{1,2}$, Hani J. Marcus$^{1,3}$, Sophia Bano$^{1,2}$}
\IEEEauthorblockA{$^{1}$UCL Hawkes Institute, University College London, London, UK\\
$^{2}$Department of Computer Science, University College London, London, UK\\
$^{3}$Department of Neurosurgery, National Hospital for Neurology and Neurosurgery, London, UK\\
E-mail: a.wijekoon@ucl.ac.uk\\}
}

\maketitle

\begin{abstract}
Accurate intra-operative Remaining Surgery Duration (RSD) predictions allow for anaesthetists to more accurately decide when to administer anaesthetic agents and drugs, as well as to notify hospital staff to send in the next patient. Therefore RSD plays an important role in improving patient care and minimising surgical theatre costs via efficient scheduling.
In endoscopic pituitary surgery, it is uniquely challenging due to variable workflow sequences with a selection of optional steps contributing to high variability in surgery duration. 
This paper presents PitRSDNet for predicting RSD during pituitary surgery, a spatio-temporal neural network model that learns from historical data focusing on workflow sequences. PitRSDNet integrates workflow knowledge into RSD prediction in two forms: 1) multi-task learning for concurrently predicting step and RSD; and 2) incorporating prior steps as context in temporal learning and inference. PitRSDNet is trained and evaluated on a new endoscopic pituitary surgery dataset with 88 videos to show competitive performance improvements over previous statistical and machine learning methods. The findings also highlight how PitRSDNet improve RSD precision on outlier cases utilising the knowledge of prior steps.
\end{abstract}

\section{Introduction}
\label{sec:intro}


Per minute, a surgical theatre costs approximately \textdollar36~ in California, USA (in 2018) ~\cite{childers2018understanding} and \textsterling16~ in the UK (in 2011) ~\cite{abbott2011factors}. One way to reduce this cost is effective theatre scheduling via minimising idle time during surgery ~\cite{childers2018understanding, dexter2005operating}. Intra-operative \gls{RSD} prediction has been identified as a key contributor to effective workflow planning and theatre scheduling~\cite{jiao2022continuous, bellini2024artificial}. During surgery, the anaesthesia team is responsible for the safety and comfort of the patient. They monitor and interpret clinical signs of pain and/or depth of anaesthesia to adjust medicines, breathing, temperature, fluids and blood pressure. \gls{RSD} assists anaesthetics in their decision-making towards reducing the time under anaesthesia and on mechanical ventilation, potentially improving patient recovery and reducing postoperative complications~\cite{aksamentov2017deep,twinanda2018rsdnet}. However, manual \gls{RSD} prediction is difficult due to the variability of individual operations \cite{Travis2014}, and so automated techniques may provide more accurate and reliable predictions.

\begin{figure}[!t]
\centering
\includegraphics[width=.44\textwidth]{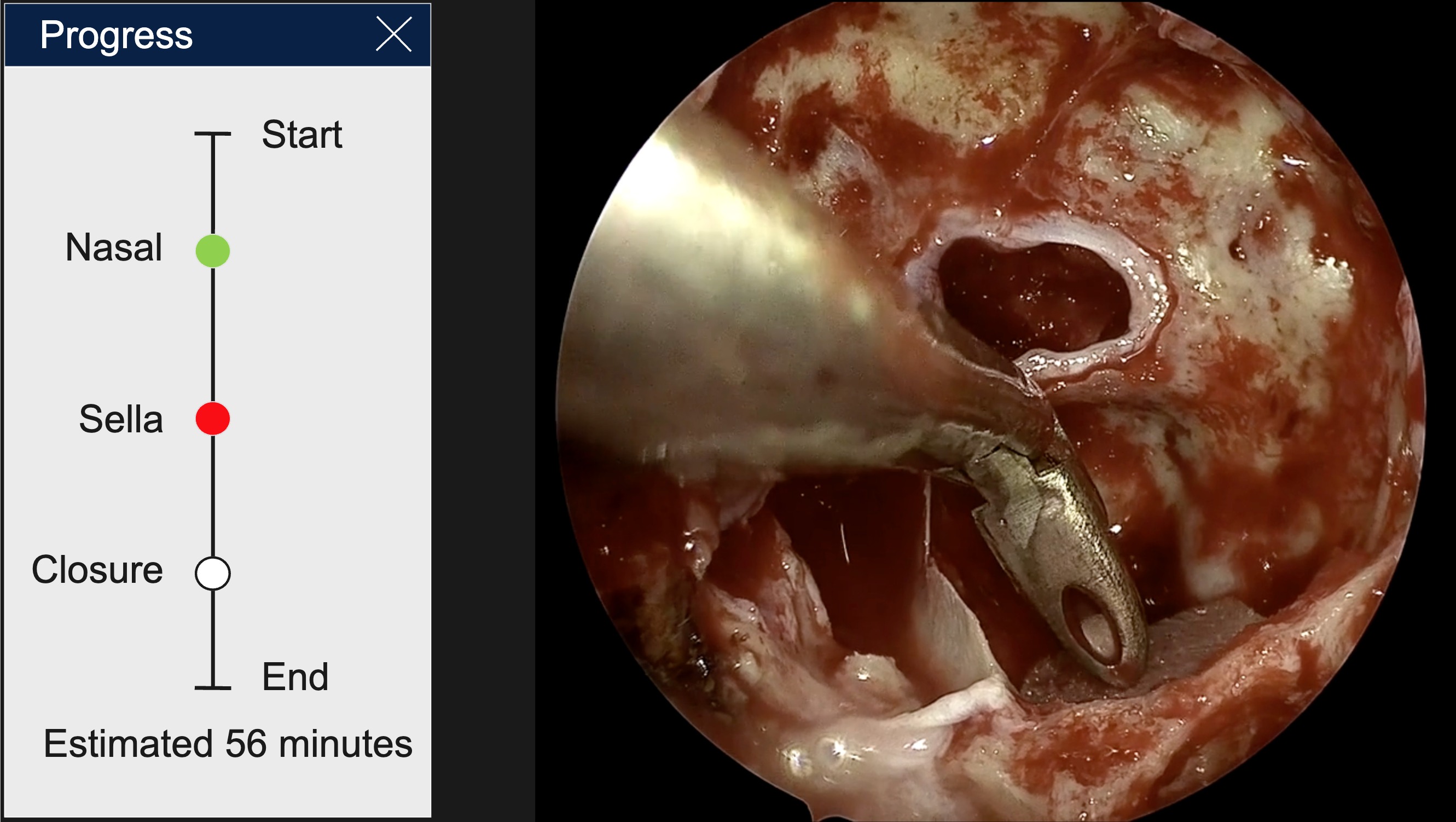}
\caption{Prototype with step progress indicators and estimated time remaining in the surgery}
\label{fig:mock}
\end{figure}

Pituitary adenomas, benign tumours of the pituitary gland, are common and often associated with systemic health issues and increased mortality \cite{Khan2023}. Most of these tumours can be effectively treated with the \gls{eTSA}, a minimally invasive surgery that removes these tumours via a nostril \cite{Marcus2021}. Recent developments in \gls{ML} have allowed for automated intra-operative decision support systems in the form of step recognition~\cite{khan2022workflow, das2022reducing,das2023automatic} and critical anatomical identification~\cite{das2023multi,mao2024pitsurgrt}. 
Extensions of these models provide opportunities to support the wider surgical~\cite{khan2024video1,khan2024video2} and non-surgical teams including anaesthetists, theatre nurses and theatre managers - such as in the form of a progress bar as presented in Figure~\ref{fig:mock}.


In \gls{eTSA}, anaesthetist find intra-operative \gls{RSD} predictions to be informative in the following events: (a) when to titrate down anaesthetic agents and allow time to wear off for a prompt wake-up; (b) when to administer pain relief and antiemetic drugs for post-operative recovery; and (c) when to notify scheduling staff to send the next patient. Considering the washout time of modern anaesthetic agents, these predictions are most useful in the last 10-20 minutes of the surgery. Accordingly, a clinically appropriate \gls{RSD} prediction model is expected to have an error of less than 5 minutes in the last 10-20 minutes and an error of less than 10 minutes over the full duration of the surgery.

Previous intra-operative \gls{RSD} estimation methods include both statistical and \gls{ML} approaches. Statistical approaches exploit duration statistics and workflow annotations from historical data to derive \gls{RSD} estimations. More recent \gls{ML} methods employ spatio-temporal neural architectures to learn \gls{RSD} prediction from historical video data. CataNet architecture for cataract surgery utilises steps and surgeon's experience as additional context to guide \gls{RSD} prediction~\cite{marafioti2021catanet}. Conversely, RSDNet proposed for cholecystectomy and gastric bypass procedures predicts \gls{RSD} without requiring additional context and hence eliminates the need for expert annotations~\cite{twinanda2018rsdnet}. Both approaches utilised a supervised multi-task approach to train their models. 

Pituitary surgeries present unique challenges when predicting \gls{RSD} stemming from its workflow complexity indicated by: 1) the diverse sequences in which surgical steps are performed; 2) the inclusion of optional steps; and 3) historical data indicating significant variability in surgery duration. To overcome these challenges, this paper introduces PitRSDNet, a \gls{ML} model that integrates previous step predictions when learning temporal dependencies in a spatio-temporal neural architecture for improved performance over previous work. The contributions are therefore as follows:

\begin{itemize}
    \item The introduction of PitRSDNet, a neural network regression model capable of accurately predicting the remaining time of endoscopic pituitary surgery.
    \item A thorough comparison of PitRSDNet and existing \gls{RSD} models on a new pituitary surgery dataset containing 88 videos.
\end{itemize}

\section{Methods}
\label{sec:method}
\subsection{Problem formulation}
\gls{RSD} prediction is approached through statistical modelling or as a \gls{ML} regression problem. Consider a surgical video where the full duration is $T$ and at a given timestamp $t$ elapsed time $t_{el} = t$, hence the remaining surgical duration is $t_{rsd} = T - t_{el}$ and progress is defined as $p = t_{el}/T$. During the surgery, $T$ is not yet known. Accordingly, statistical modelling computes $t_{rsd}$ from a reference full duration $T_{ref}$ derived from the contextual information of the ongoing surgery and historical video data, whereas a \gls{ML} regression model predicts $t_{rsd}$ based on the input and its learned parameters. 

\subsection{Proposed PitRSDNet}

\begin{figure}[!t]
    \centering
    \includegraphics[width=.49\textwidth]{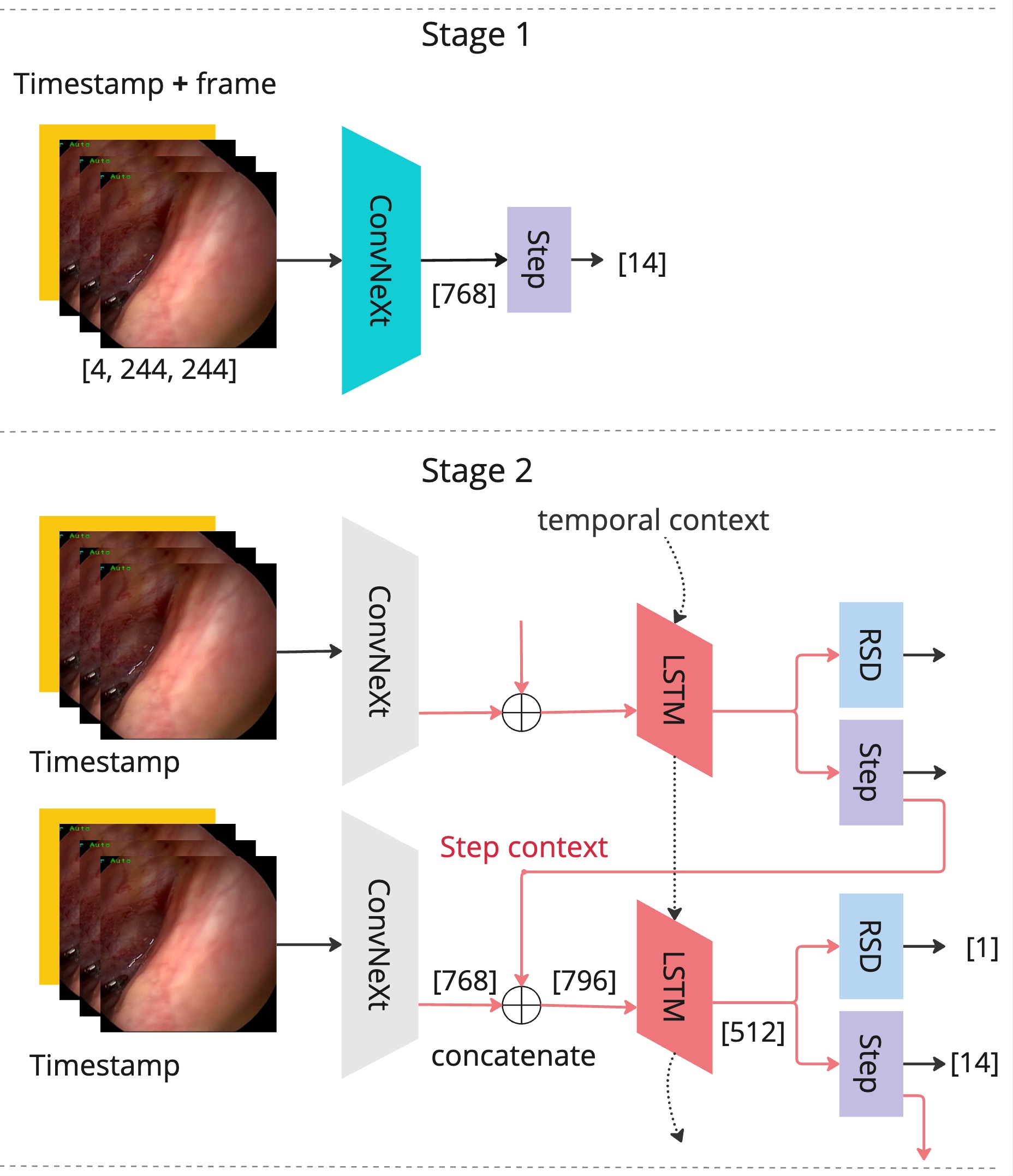}
    \caption{PitRSDNet architecture and training stages}
    \label{fig:arch}
\end{figure}

PitRSDNet is a multi-task deep neural architecture trained in two stages that incorporate step transition knowledge into \gls{RSD} prediction.
The input to the model at timestamp $t$ is $I_t \in \mathbb{R}^{4\times244\times244} $ where the timestamp is considered an additional channel similar to CataNet~\cite{marafioti2021catanet}. The proposed architecture and the training stages are presented in Figure~\ref{fig:arch} and are presented below in detail. 

At the first stage, a pre-trained ConvNeXt~\cite{liu2022convnet} encoder, $f(.)$, is fine-tuned for step classification. The weight average technique from ~\cite{marafioti2021catanet} is followed to adapt the input layer weights for the 4-channel input. ConvNeXt is selected as the encoder over ResNet~\cite{he2016deep,twinanda2018rsdnet} and DenseNet~\cite{huang2017densely,marafioti2021catanet} used in previous \gls{RSD} models. This is motivated by its superior performance over other convolutional and transformer architectures in the public domain~\cite{liu2022convnet} and for its use of Layer Normalisation to mitigate information leaks during online tasks~\cite{rivoir2024pitfalls}. Weighted cross-entropy is used in fine-tuning to account for the class imbalance in step labels.

The second stage trains two \gls{LSTM} layers with two output heads, $g(.)$, for \gls{RSD} and step prediction. To improve step prediction performance, inspired by~\cite{ban2021aggregating} and~\cite{rivoir2024pitfalls}, context from previous step predictions into the \gls{LSTM} input, $l_t$, is incorporated. Accordingly, $l_t$ is formed by concatenating the frozen fine-tuned ConvNeXt encoder output of the input, step prediction probabilities for the last frame, and mean step prediction probabilities for the last $\hat{t}$ frames~(see Equation~\ref{eq:step_context}). 
The training is guided by an unweighted sum of weighted cross-entropy loss from step and Smooth L1 loss from \gls{RSD} prediction errors. 

PitRSDNet adapts the \gls{RSD} normalisation proposed in RSDNet to regularise model training~\cite{twinanda2018rsdnet}. Considering pituitary surgery duration (minutes), we select a normalisation factor of 10 for all RSD ground-truth values and predictions. 

\begin{equation}
\begin{gathered}
l_t = f(I_t) \oplus s_{t-1}\\
s_{t-1} = g(l_{t-1})^{step} \oplus \frac{1}{\hat{t}} \sum_{i=1}^{\hat{t}} g(l_{t-i})^{step} 
\end{gathered}
\label{eq:step_context}
\end{equation}

\subsection{Baseline methods}

In evaluation, we aim to compare the performance of PitRSDNet against the following statistical and \gls{ML} baselines.

\begin{itemize}
\item \textbf{Naive} approach~\cite{aksamentov2017deep} calculated the reference full duration, $T_{ref}$ from historical data to derive $t_{rsd} = max(0, T_{ref} - t_{el})$. We use the mean full duration of the training split as $T_{ref}$ (for each respective dataset).

\item \textbf{Step-inferred} approach~\cite{aksamentov2017deep} utilised workflow annotations to derive $t_{rsd}$. Consider the sequence of $N$ surgical steps, where $t^s_{ref}$ is the reference duration of step $s$. The $t_{rsd}$ is calculated considering the time elapsed at the current step $s$ and the reference durations of remaining steps as $t_{rsd} = max(0, t^s_{ref} - t^s_{el}) + \sum_{i=s+1}^{N} t^i_{ref}$. This approach assumes that a surgical workflow consists of a set of mandatory steps that occur sequentially. Similar to \Q{Naive}, the mean of the training split is used to calculate the reference step durations. 

\item \textbf{Sequence matching} approach is proposed to address the non-sequential nature of pituitary surgical steps. The sequence of elapsed steps is compared with the sub-sequences from historical videos up to the same timestamp using Levenshtein distance. This yields a similarity score between the current video and each historical video, identifying the most similar ones. The reference full duration $T_{ref}$ is calculated as the mean full duration from the top $k$-nearest neighbours $T_{ref} = \frac{1}{k}\sum_{i=0}^k (T^i)$ and is used to calculate the $t_{rsd}$ as $max(0, T_{ref} - t_{el})$. For pituitary datasets, k=3 and sequences are compressed to reduce the computational complexity. For Pit-33, which includes instrument annotations, the elapsed instrument sequence is matched against historical data. The similarity to a historical video is then calculated as the unweighted sum of step and instrument sequence similarities.

\item \textbf{RSDNet}~\cite{twinanda2018rsdnet} considered \gls{RSD} prediction as a \gls{ML} task. RSDNet consists of a ResNet~\cite{he2016deep} backbone and an \gls{LSTM} layer where the input is video frames. RSDNet incorporates elapsed time to the \gls{LSTM} output and predicts \gls{RSD} and progress in a multi-task manner. The model training involves two stages, for further details, refer to~\cite{twinanda2018rsdnet}. 

\item \textbf{CataNet}~\cite{marafioti2021catanet} proposed a spatio-temporal neural architecture that incorporates workflow annotations to predict \gls{RSD}. Their DenseNet backbone learned spatial dependencies from frames where the elapsed timestamp forms one of the four input channels. The training involves 4 stages, see details in~\cite{marafioti2021catanet}.
Originally CataNet utilised a binary classification head to predict the experience of the surgeon~(1: experienced; 0: novice). This is guided by the knowledge that novice surgeons take longer to complete a surgery. This did not translate to pituitary surgery, accordingly, we excluded the experience prediction head.

\end{itemize}

In evaluation, we also consider the following ablated variants of the PitRSDNet architecture to assess the impact of prior step context integration and the necessity of workflow annotations for \gls{RSD} prediction in pituitary surgery.
\begin{itemize}
\item \textbf{PitRSDNet(RSD)}: only stage 2 trained for RSD prediction, no step prediction head or prior step context integration
\item \textbf{PitRSDNet(S,RSD)}: PitRSDNet without prior step context integration
\item \textbf{PitRSDNet(S,I,RSD)}: PitRSDNet with an additional instrument prediction head in both stages; no prior step context integration and only applicable for Pit-33.
\end{itemize}

\section{Dataset Description}
\label{sec:data}

For experimental evaluation, two datasets are utilised, namely, Pit-88 and Pit-33: (i) Pit-88 consists of 88-videos with step annotations; (ii) Pit-33, a subset of Pit-88, consists of 33-videos with both step and instrument annotations. 25-videos of Pit-33 are publicly available~\cite{das2024}, as presented in PitVis-EndoVis MICCAI-2023 sub-challenge \cite{EndoVis2023}.  Table~\ref{tbl:datasets} summarises both datasets. 

\subsection{Images}
All videos were collected from two consultant surgeons at a single-centre (National Hospital of Neurology and Neurosurgery, London, United Kingdom) between 2018 and 2023 with informed patient consent. Ethical approval was granted for the project via the \gls{IRB} at \gls{UCL} (17819/011), and informed participation consent was obtained. A high-definition endoscope (Hopkins Telescope, Karl Storz Endoscopy) was used to record the surgeries. 
All videos were uploaded and analysed using Touch Surgery™ Ecosystem, an AI-powered surgical video management and analytics platform provided by Medtronic~\footnote{https://www.touchsurgery.com/}.
Using their internal software, all images outside of the patient were blurred to de-identify the patient. The videos were then reduced to 720p (1280 $\times$ 720) resolution at 24-\gls{FPS} using the publicly available software handbrake\footnote{https://handbrake.fr/}, and stored as mp4 files. Images were sampled from the videos at 1-\gls{FPS}; centre cropped to 720 $\times$ 720 to remove the excessive black borders; resized to 256 $\times$ 256; and stored as .png.
\begin{figure*}[!t]
\centering
\subfigure[Pit-33]{\includegraphics[width=.45\textwidth]{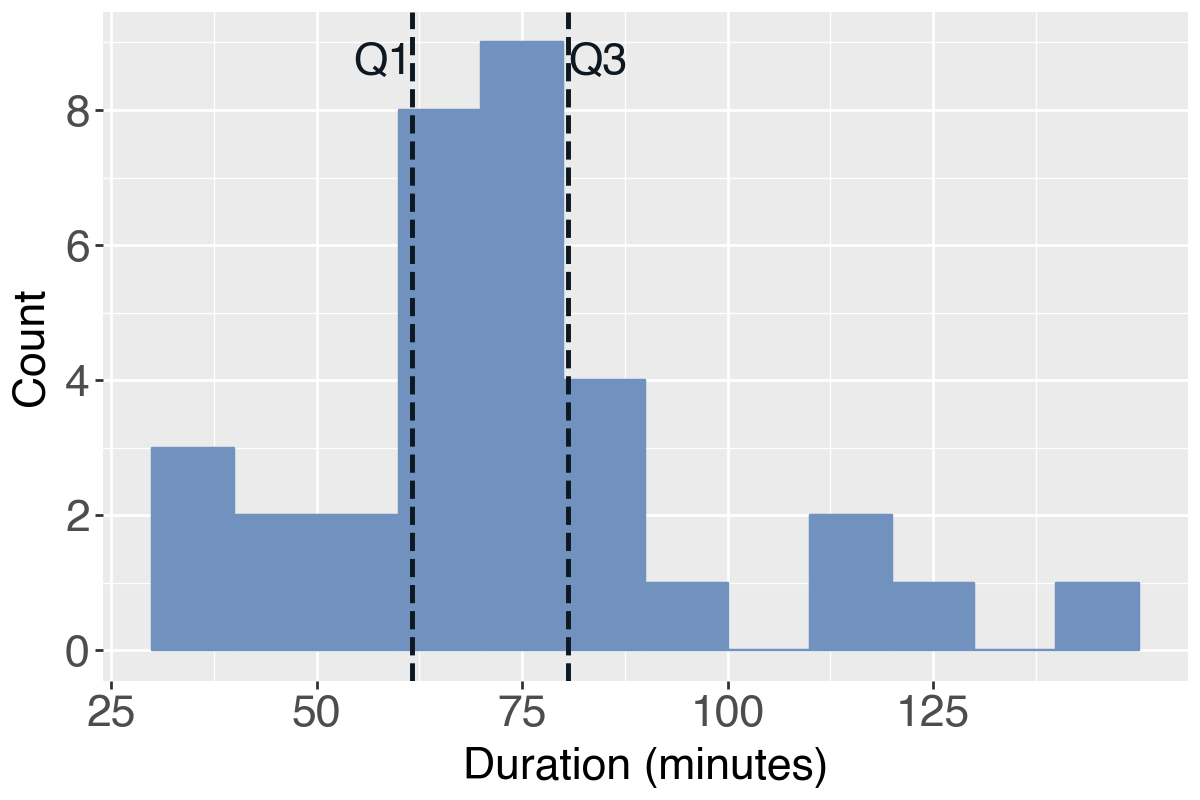}}
\subfigure[Pit-88]{\includegraphics[width=.45\textwidth]{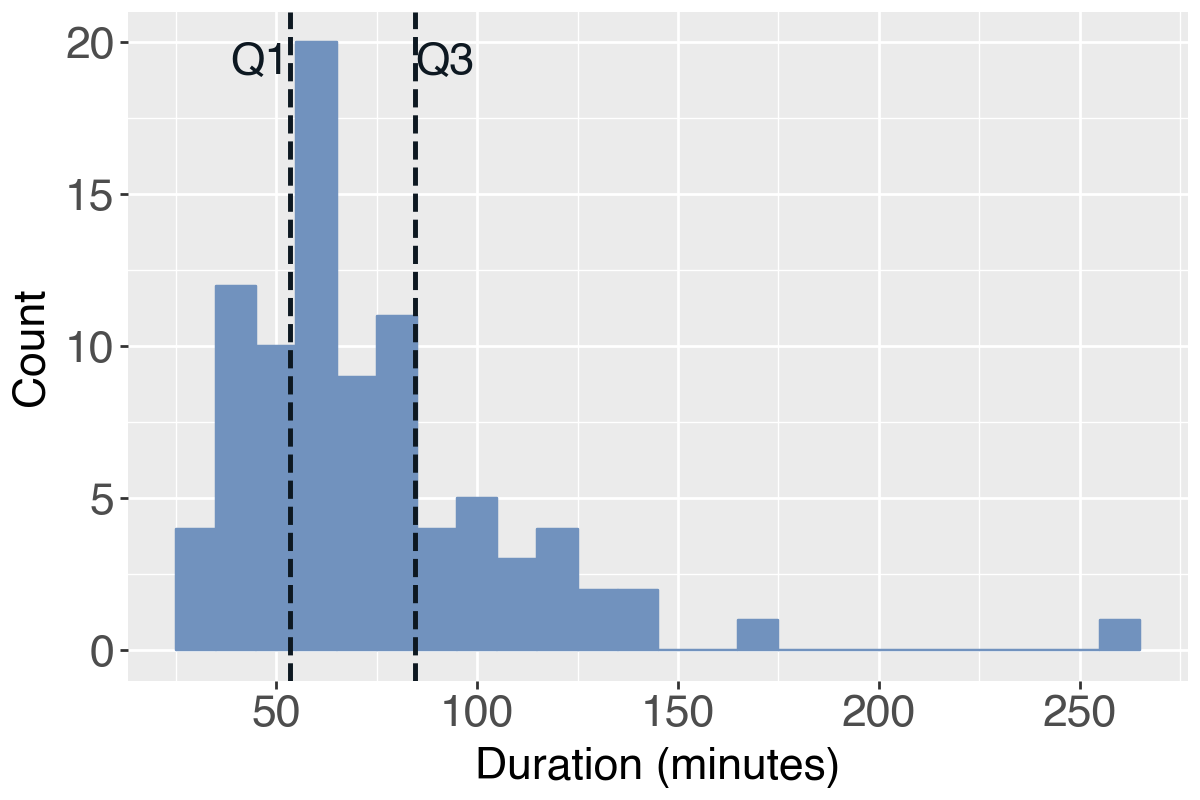}}
\caption{Surgery duration distributions in (a) Pit-33 and (b) Pit-88 annotated with 25th~(Q1) and the 75th~(Q3) percentiles. For Pit-33, the median duration is 72 minutes (inter-quartile range (IQR): 61-80 minutes), and for Pit-88, it is 64 minutes (IQR: 53-84 minutes)}
\label{fig:dist-duration}
\end{figure*}

\begin{figure}[!ht]
\centering
{\includegraphics[width=0.46\textwidth]{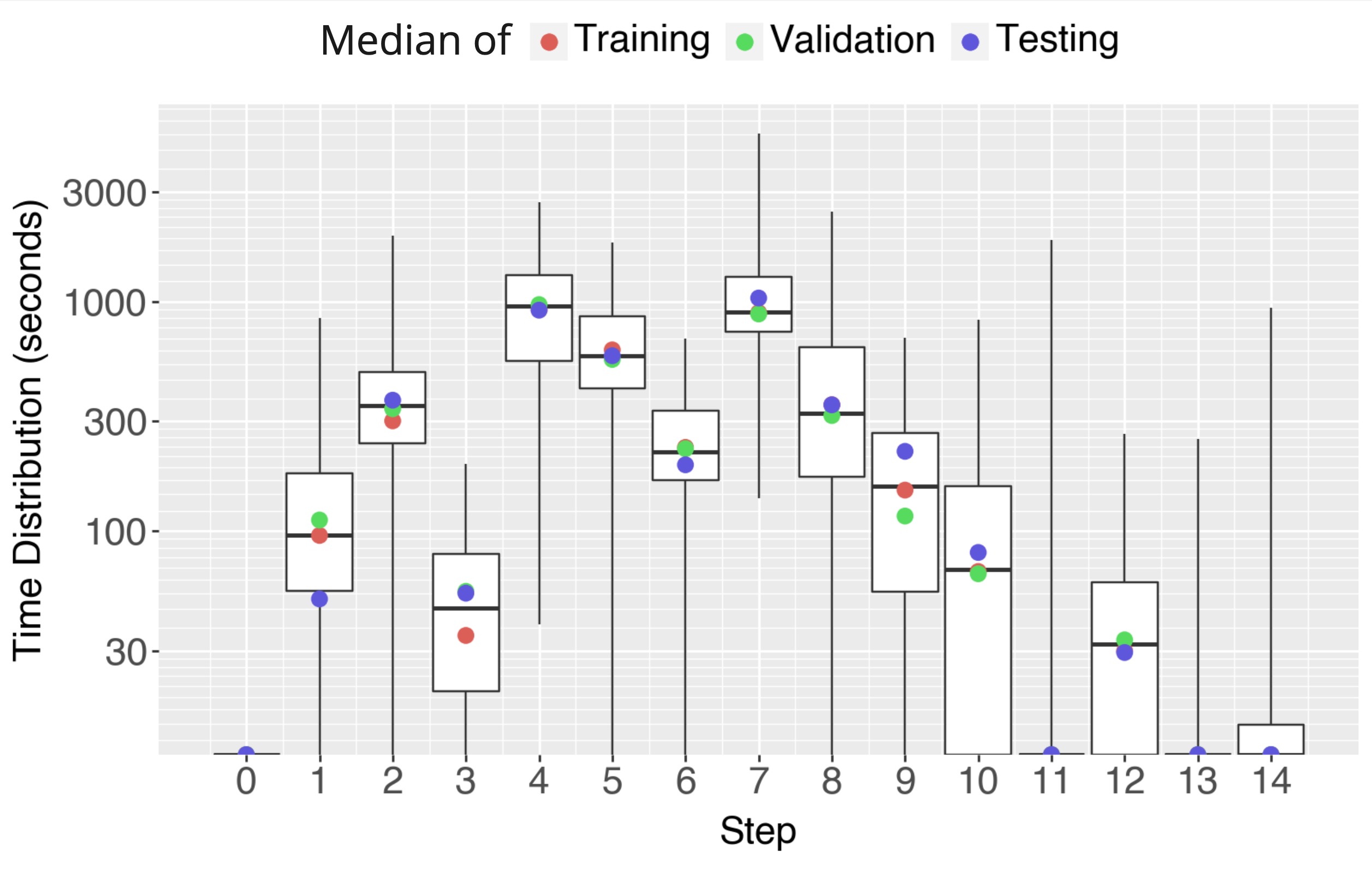}}
\caption{Distribution of time in seconds associated with each step in the Pit-88 dataset. Points refer to the median of training, validation and testing splits.}
\label{fig:step-dist}
\end{figure}

\begin{table}[!ht]
\centering
\caption{Dataset statistics overview of the two datasets used.}
\label{tbl:datasets}
\begin{tabular}{lrr}
\hline
&Pit-33&Pit-88\\
\hline
Number of surgeries &33&88\\
Train/val/test splits&20/5/8&70/8/10\\
Duration total (minutes)&74.69$\pm$23.63&73.62$\pm$35.43\\
Train duration (minutes)&72.68$\pm$20.97&76.18$\pm$41.57\\
Val duration (minutes)&101.12$\pm$23.77&71.55$\pm$33.72\\
Test duration (minutes)&63.23$\pm$19.37&71.79$\pm$27.92\\
Number of steps&15&14\\
Number of instruments&19&N/A\\
\hline
\end{tabular}
\end{table}

\subsection{Annotations}
The steps annotations correspond to 14 surgical steps. with 8-core and 6-optional steps. Additionally, these steps are not necessarily performed in sequence and the same step may appear multiple times during the surgery (e.g. Step-8 \Q{haemostasis} can be performed at any stage of the surgery considering the level of bleeding). Pit-33 step annotations include an additional \Q{out of patient} step corresponding to the de-identified images. For the 55-videos in Pit-88 where this annotation is not available, de-identified frames are assumed to be part of the previous step.

Instrument annotations of Pit-33 consider 18 instruments. Each video contained between 9 to 18 surgical instruments, with all 18 instruments appearing in only 5 out of the 33 videos.

\subsection{Duration distribution}

Figure~\ref{fig:dist-duration} presents the distribution of surgery duration annotated with 25th~(Q1) and the 75th~(Q3) percentiles. From this, it can be seen most videos fall between 60-80-minutes, with some extreme outliers with much shorter (< 40-minutes) or much longer (> 125-minutes) durations. 
Furthermore, Figure~\ref{fig:step-dist} presents the step distributions of Pit-88, where it can be seen there is quite a large variation in step length, which is important to consider in \gls{RSD} calculations. 

From Table~\ref{tbl:datasets} it can be seen Pit-88 achieves comparable distributions across the 3 splits whereas Pit-33 validation distribution is significantly different from the training and testing splits. This may affect results as many statistical \gls{RSD} models rely on having a similar video to compare the video they are attempting to predict the \gls{RSD} for.

\section{Evaluation}

\subsection{Implementation details}

Naive and Step-inferred methods were implemented following the details provided by the authors of~\cite{twinanda2018rsdnet}. The Levenshtein distance for the sequence matching approach was implemented using the Python library textdistance\footnote{https://github.com/life4/textdistance}. 

For \gls{ML} methods, the dataset split is given in Table~\ref{tbl:datasets}. Horizontal flip, crop, and resize to 224$\times$224 augmentations were applied at random in the data loader. The code was written in PyTorch 2.2 with training done on a 32GB NVIDIA Tesla V100 Tensor Core GPU.  
RSDNet and CataNet models were adapted for pituitary datasets from the GitHub repository provided by the authors of~\cite{marafioti2021catanet}. Both adaptations reused the implementation and training hyper-parameter provided in respective publications. PitRSDNet code was adapted from the same GitHub repository. ConvNeXt was fine-tuned for 3 epochs with a learning rate of 0.0001; multi-task \gls{LSTM} was trained for 40 epochs with a learning rate starting at 0.001 for the first 20 epochs and reduced to 0.0001 for the last 20. During \gls{LSTM} training \gls{RSD} normalisation resulted in a 2:3 loss ratio between step~(and instruments when available) and \gls{RSD} losses.

\begin{table*}[t]
\centering
\caption{Performance achieved by Statistical Methods reported by \gls{MAE} 1) in the last 10 minutes; 2) in the last 20 minutes; and 3) over the full duration. Bold values indicate the best performance achieved and the $\dagger$ symbol indicates statistical significance over the next best-performing method. }
\label{tbl:naive}
\begin{tabular}{llrrr}
\hline
\multirow{2}{*}{Dataset}&\multirow{2}{*}{Method}&\multicolumn{3}{c}{Mean Absolute Error (minutes)$\downarrow$}\\
&&Last 10&Last 20&Full Duration\\
\hline
\multirow{4}{*}{Pit-33}&Naive (Mean)&47.99$\pm$35.63&45.64$\pm$34.15&31.28$\pm$17.35\\
&Step-inferred (Mean)&46.75$\pm$25.32&41.56$\pm$24.31&27.90$\pm$9.45\\
&Sequence matching (S)&\textbf{8.20$\pm$4.41}&\textbf{11.97$\pm$6.94}&15.25$\pm$7.40\\
&Sequence matching (S,I)&14.19$\pm$15.38&14.05$\pm$13.33&\textbf{14.71$\pm$10.99}\\
\hline
\multirow{3}{*}{Pit-88}&Naive (Mean)&61.54$\pm$38.38&59.04$\pm$39.19&40.86$\pm$31.56\\
&Step-inferred (Mean)&56.15$\pm$33.46&48.17$\pm$34.21&32.30$\pm$19.96\\
&Sequence matching (S)&\textbf{5.75$\pm$3.71}$\dagger$&\textbf{8.52$\pm$3.93}$\dagger$&\textbf{16.69$\pm$14.98}\\
\hline
\end{tabular}
\end{table*}

\begin{table*}[t]
\centering
\caption{Performance achieved by Machine Learning Methods reported by \gls{MAE} 1) in the last 10 minutes; 2) in the last 20 minutes; and 3) over the full duration. Bold values indicate the best performance achieved.}
\label{tbl:dl}
\begin{tabular}{llrrrr}
\hline
\multirow{2}{*}{Dataset}&\multirow{2}{*}{Method}&\multicolumn{3}{c}{Mean Absolute Error (minutes)$\downarrow$}&\multirow{2}{*}{Step F1-macro}\\
&&Last 10&Last 20&Full Duration&\\
\hline
\multirow{6}{*}{Pit-33}&RSDNet&10.51$\pm$9.32&11.76$\pm$8.95&15.78$\pm$7.94&N/A\\
&CataNet&13.99$\pm$15.71&13.69$\pm$15.14&14.75$\pm$12.58&0.3981$\pm$0.10\\
&PitRSDNet(RSD)&11.23$\pm$16.20&11.38$\pm$14.70&\textbf{13.07$\pm$11.63}&N/A\\
&PitRSDNet(S,RSD)&\textbf{7.87$\pm$7.84}&\textbf{9.01$\pm$8.49}&13.45$\pm$7.57&0.4191$\pm$0.09\\
&PitRSDNet(S,I,RSD)&11.87$\pm$13.07&11.21$\pm$11.40&14.96$\pm$9.36&0.4432$\pm$0.06\\
&PitRSDNet&11.15$\pm$10.60&11.23$\pm$10.93&15.53$\pm$10.14&\textbf{0.4622$\pm$0.07}\\
\hline
\multirow{5}{*}{Pit-88}&RSDNet&9.06$\pm$7.60&9.62$\pm$7.20&13.83$\pm$10.70&N/A\\
&CataNet&6.79$\pm$10.65&8.12$\pm$10.56&14.28$\pm$10.05&0.6069$\pm$0.09\\
&PitRSDNet(RSD)&9.67$\pm$12.34&9.56$\pm$0.13&16.08$\pm$11.87&N/A\\
&PitRSDNet(S,RSD)&4.33$\pm$2.54&\textbf{4.98$\pm$3.29}&13.05$\pm$10.60&0.6027$\pm$0.09\\
&PitRSDNet&\textbf{4.08$\pm$3.01}&6.20$\pm$2.87&\textbf{12.25$\pm$6.53}&\textbf{0.6361$\pm$0.10}\\
\hline
\end{tabular}
\end{table*}

\subsection{Evaluation metrics}

RSD prediction performance is measured using \gls{MAE}. First \gls{MAE} is calculated individually for each video and mean-averaged over all videos to present the final \gls{MAE}. Based on clinical motivation and previous literature \gls{MAE} is presented for the last 20 and 10 minutes of the surgeries. When comparing \gls{MAE} between methods for statistical significance paired Wilcoxon signed-rank test accounting for the varied lengths of the surgery videos was used. 

Step prediction performance of CataNet and PitRSDNet is measured using macro-F1 score accounting for the class imbalance. While step prediction is not the focus of this study, we compare the step performance with the state-of-the-art performance published in~\cite{das2022reducing} and~\cite{das2023automatic}.

\section{Results and Discussion}
\label{sec:eval}

\begin{figure*}[!h]
\centering
\includegraphics[width=1\textwidth]{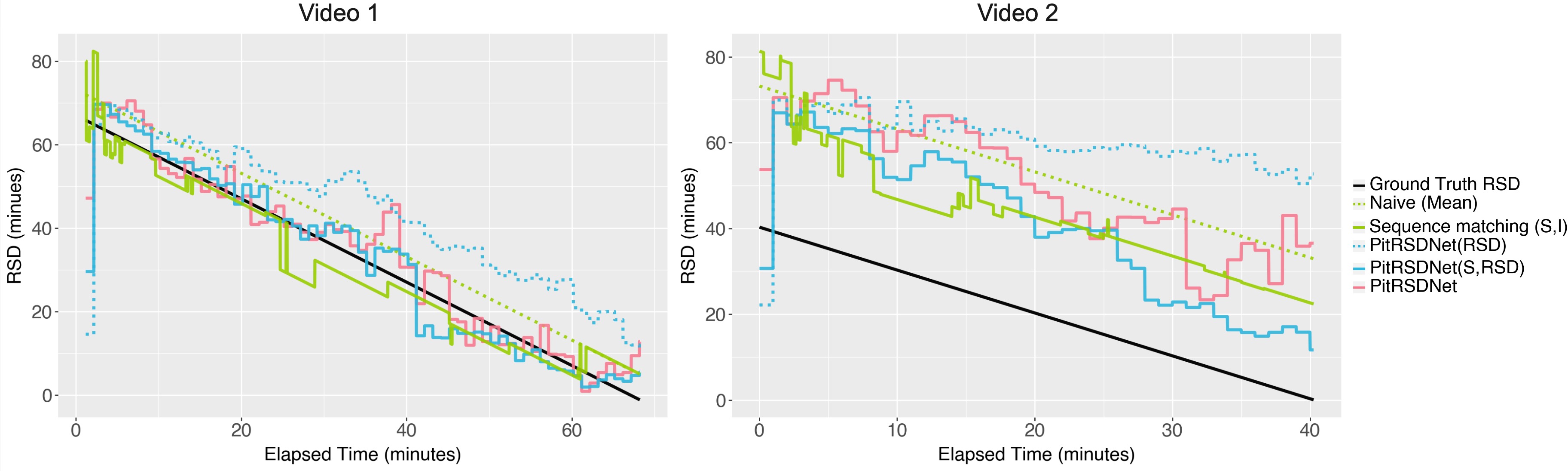}
\caption{RSD predictions over the surgery duration of two videos in Pit-33 test set. Video 1 duration closely resembles training data while video 2 is significantly shorter in duration - see ground truth RSD indicated in Black. Other lines refer to RSD prediction methods - see legend.}
\label{fig:33-vis}
\end{figure*}
\begin{figure*}[!h]
\centering
\includegraphics[width=1\textwidth]{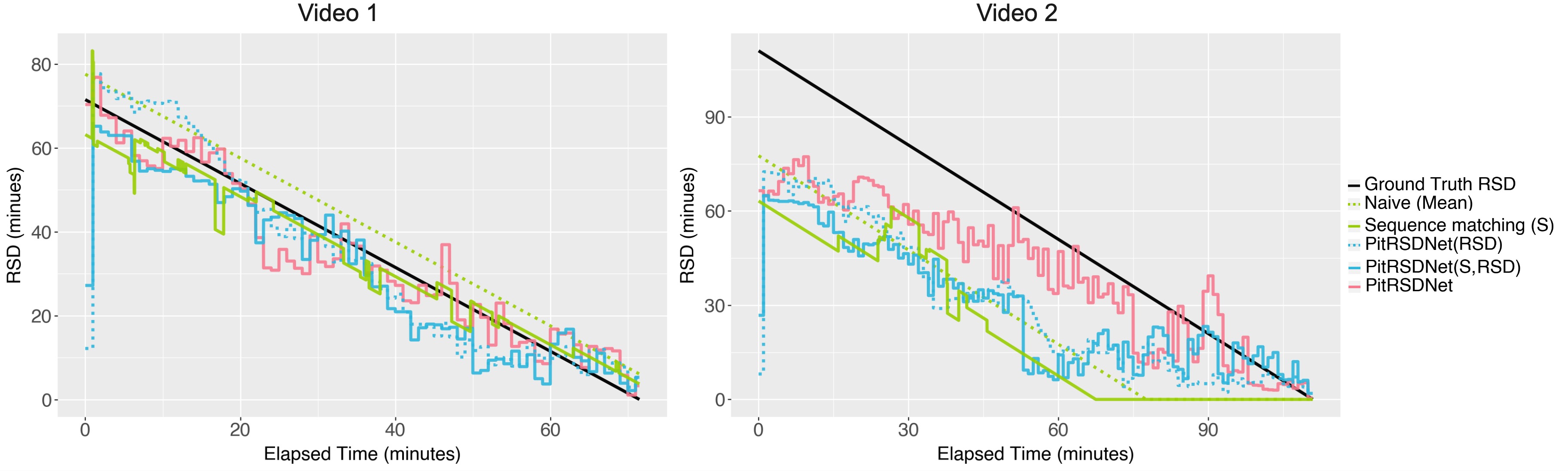}
\caption{RSD predictions over the surgery duration of two videos in Pit-88. Video 1 duration closely resembles training data while video 2 is significantly longer in duration - see ground truth RSD indicated in Black. Other lines refer to RSD prediction methods - see legend. }
\label{fig:88-vis}
\end{figure*}

\subsection{Quantitative results} Tables~\ref{tbl:naive} and~\ref{tbl:dl} present the performances achieved by statistical and \gls{ML} methods, respectively. Bold values indicate the best performance achieved and $\dagger$ symbol indicates statistical significance over the next best-performing method. 

Results in Table~\ref{tbl:naive} show that sequence matching methods significantly outperform both Naive and Step-inferred methods~\cite{aksamentov2017deep}. Sequence matching on steps reduced the mean error by 12.01 and 13.83 minutes over the Step-inferred method, despite both methods utilising the same contextual information from historical data. 
Pit-33 utilisation of instrument labels has further reduced the full duration error, however, it has been detrimental in the last 20 minutes of the surgery.

Table~\ref{tbl:dl} shows that either PitRSDNet or its ablated variants achieve improved performance over RSDNet and CataNet. PitRSDNet(RSD) archived the best performance with Pit-33 over the full duration, whereas including the step head~(PitRSDNet(S,RSD)) has significantly improved the performance during the last 20 minutes. 
Similar to Non-ML methods, adding the instrument head with Pit-33 has been detrimental to \gls{RSD} performance however, it has a significant impact on the step prediction performance. 

With Pit-33, PitRSDNet did not achieve competitive \gls{RSD} performance, however, it significantly improves step recognition~($\sim4.3\%$ over PitRSDNet(S,RSD)). 
Conversely, with Pit-88, PitRSDNet achieves the best performance in both \gls{RSD} and step prediction. This lack of performance is likely to do with the smaller Pit-33 dataset size. As described in Section~\ref{sec:data}, there is a high variability in step sequences with the inclusion of \Q{out-of-patient} step. Accordingly, there is a larger error in step prediction and consequently, the prior step context is adding noise to the LSTM input. PitRSDNet minimises this error in the presence of substantial data, and the proposed architectural changes achieve competitive performance improvements. 

\subsection{Qualitative results} 
\label{sec:quali}
Figures~\ref{fig:33-vis} and~\ref{fig:88-vis} display how \gls{RSD} predictions from different methods change throughout the surgery.
In each dataset, video 1 is selected to closely align with respective training data, whereas video 2 is under-represented~(see Naive approach - dotted Green). Pit-33 video 2 is a significantly shorter surgery while Pit-88 video 2 is longer.
In each sub-figure, lines indicate RSD predictions and the ground-truth RSD is indicated in Black colour. 
For clarity, we compare PitRSDNet RSD prediction with the best~(or next best) and worst performing approaches~(on full duration \gls{MAE}) from Tables~\ref{tbl:naive} and~\ref{tbl:dl}. 
RSD predictions from ML methods are visualised at one-minute intervals, even though they predict RSD for each frame, to reflect the less frequent practical updates.

For both representative cases, sequence matching~(solid Green) closely follows the ground truth, making necessary adjustments as steps~(and instruments) change. Similarly, both PitRSDNet(S,RSD) and PitRSDNet closely resemble ground truth in these cases. 
PitRSDNet(RSD) (dotted Blue), performed poorly in the last 20 minutes with Pit-33 videos, attributed to insufficient guidance from the step head and prior step context. 
This is also reflected in Table~\ref{tbl:dl} where PitRSDNet(RSD) performed best over the full duration that is not reflected in the last 20 minutes~($\sim$3.4\% increase in error compared to PitRSDNet(S,RSD)). 

Most methods struggle with under-represented cases. For the shorter surgery (Pit-33 video 2), the lowest \gls{MAE} is 40.60$\pm$17.26 minutes per frame~(solid Blue) achieved by PitRSDNet(S,RSD). 
However, as time progresses in the longer surgery~(Pit-88 video 2), PitRSDNet~(solid Red) is making necessary corrections at the earliest utilising prior step context. This is evident in comparison to PitRSDNet(S,RSD) (solid Blue) with no prior step context, which makes similar corrections much later in the surgery. 

Considering both quantitative and qualitative results we highlight the following findings and conclusions. 
\begin{itemize}
\item Methods such as PitRSDNet and sequence matching that are informed by the clinical knowledge of pituitary surgeries achieve improved RSD prediction. 
\item All observations strongly support the utilisation of workflow annotations to improve RSD, both in multi-task learning and by learning with prior step context, as we have proposed in PitRSDNet.
\item Availability of step annotations for historical data and incorporating step context in training has improved RSD prediction. However, the best macro F1-score achieved by the step head was 0.6361$\pm$0.10 which is below the performance reported on similar datasets in~\cite{das2022reducing} and~\cite{das2023automatic}. As such in the future, we will explore more effective approaches to improving step predictions within the RSD models.
\item All methods find predicting RSD for under-represented surgeries challenging, even when the training/testing set distributions are matched. This preludes to other contributing factors~(e.g. pre-operative parameters) that need to be integrated into the RSD prediction in the future. 
\item 
In the introduction we discussed the clinically appropriate benchmarks for \gls{RSD} prediction in pituitary surgery. The results in Table~\ref{tbl:dl} showed that PitRSDNet achieves the clinical benchmark in the last 10-20 minutes~(i.e., error $\leq$5 minutes) and is the only method to do so. PitRSDNet is the closest but narrowly failed to meet the clinical benchmark over the full duration of the surgery~(missing by 2-3 minutes). Figure~\ref{fig:88-vis} showed that surgeries underrepresented in duration contributed to the error margins. These observations, compared with the clinical benchmarks, highlight the continued work required to improve the precision of \gls{RSD} prediction.

\item A surgeon's experience and surgical philosophy are two key factors influencing the endonasal pituitary surgical workflow and surgical duration. This is consistent with previous research in \gls{RSD} which showed that surgeon-specific modelling yields improved performance over surgeon-agnostic models~\cite{bellini2024artificial,spence2023machine}. This paper used data from two consultant neurosurgeons and our surgeon-agnostic approach to \gls{RSD} prediction achieved clinically significant performance. In the future, we will explore the transferability of this work to multiple centres and multiple experience levels as data is made available.
\end{itemize}

\section{Conclusion}
\label{sec:conc}
This paper presents PitRSDNet, a neural architecture for predicting the remaining time in endoscopic pituitary surgery. The findings highlighted how existing methods struggle to accurately predict \gls{RSD} considering the complex workflow sequences seen in pituitary surgery. Proposed changes in PitRSDNet that integrated prior steps as context resulted in improved \gls{RSD} prediction while improving the precision of step recognition. The detailed findings lead our ongoing and future work including improving the performance of outlier cases towards meeting the clinical benchmarks.

\section*{Acknowledgement}
With thanks to Digital Surgery Ltd, a Medtronic company, for access to Touch Surgery Ecosystem for video recording, annotation and storage. For the purpose of open access, the author has applied a CC BY public copyright licence to any author-accepted manuscript version arising from this submission.

\section*{Funding}
This work was supported in whole, or in part, by the \gls{WEISS} [203145/Z/16/Z], the \gls{EPSRC} [EP/W00805X/1, EP/Y01958X/1], the \gls{DSIT} and the Royal Academy of Engineering under the Chair in Emerging Technologies programme. Adrito Das is supported by \gls{EPSRC} [EP/S021612/1]. Danyal Z. Khan is supported by the \gls{NIHR} Academic Clinical Fellowship [2021-18-009] and the \gls{CRUK} Pre-doctoral Fellowship [RCCPDB-Nov21/100007]. Hani J. Marcus is supported by \gls{WEISS} [NS/A000050/1] and by the \gls{NIHR} Biomedical Research Centre at \gls{UCL}.
\bibliography{ref}

\end{document}